\documentclass{llncs}

\usepackage{amsfonts}
\usepackage{amsmath}
\usepackage{amssymb}
\usepackage{latexsym}
\usepackage{color}
\usepackage{graphics}
\usepackage{enumerate}
\usepackage{psfig}
\usepackage{amstext}
\usepackage{pseudocode}
\usepackage[whileod]{algochl}
\usepackage{gastex}
\usepackage{url}
\usepackage{ifthen}
\usepackage{times}

\def\Kset{\mathbb{K}}
\def\Nset{\mathbb{N}}

\def\Rset{\mathbb{R}}

\newcommand{\bi}{\begin{itemize}}
\newcommand{\ei}{\end{itemize}}
\newcommand{\be}{\begin{enumerate}}
\newcommand{\ee}{\end{enumerate}}
\newcommand{\bd}{\begin{description}}
\newcommand{\ed}{\end{description}}

\newcommand{\zero}{\overline{0}}
\newcommand{\one}{\overline{1}}

\newcommand{\K}{(\Kset, \oplus, \otimes, \zero, \one)}

\newcommand{\den}[1]{[\![ #1 ]\!]}
\newcommand{\set}[1]{\{#1\}}

\newcommand{\ipsfig}[2]{\scalebox{#1}{\psfig{#2}}}

\newcommand{\ignore}[1]{}

\newcommand{\e}{\epsilon}

\newcommand{\nsp}{\!\!}
\newcommand{\coln}{\nsp:\nsp}

\newcommand{\da}{\mathop{\mbox{\rm da}}}
\newcommand{\dpa}{\mathop{\mbox{\rm dpa}}}

\title{General Algorithms for Testing \\the Ambiguity of Finite Automata}

\author{
Cyril Allauzen\inst{1} \fnmsep \thanks{~~This author's new address is: Google Research, 76 Ninth Avenue, New York, NY 10011.}
\and Mehryar Mohri\inst{1, 2}
\and Ashish Rastogi\inst{1}
\institute{Courant Institute of Mathematical Sciences,\\
251 Mercer Street, New York, NY 10012.
\and
Google Research,\\
76 Ninth Avenue, New York, NY 10011.}}

\begin{document}

\maketitle

\begin{abstract}
  This paper presents efficient algorithms for testing the finite,
  polynomial, and exponential ambiguity of finite automata with
  $\e$-transitions. It gives an algorithm for testing the exponential
  ambiguity of an automaton $A$ in time $O(|A|_E^2)$, and finite or
  polynomial ambiguity in time $O(|A|_E^3)$. These complexities
  significantly improve over the previous best complexities given for
  the same problem. Furthermore, the algorithms presented are simple
  and are based on a general algorithm for the composition or
  intersection of automata. We also give an algorithm to determine the
  degree of polynomial ambiguity of a finite automaton $A$ that is
  polynomially ambiguous in time $O(|A|_E^3)$. Finally, we present an
  application of our algorithms to an approximate computation of the
  entropy of a probabilistic automaton.

\end{abstract}

\section{Introduction}

The question of the ambiguity of finite automata arises in a variety
of contexts.  In some cases, the application of an algorithm requires
an input automaton to be finitely ambiguous, in others the convergence
of a bound or guarantee relies on that finite ambiguity or the asymptotic
rate of the increase of ambiguity as a function of the string length. Thus,
in all these cases, one needs an algorithm to test the ambiguity, either
to determine if it is finite, or to estimate its asymptotic rate of increase.

The problem of testing ambiguity has been extensively analyzed in the
past. The problem of determining the degree of ambiguity of an
automaton with finite ambiguity was shown to be
PSPACE-complete. However, testing finite ambiguity can be done in
polynomial time using a characterization of polynomial and exponential
ambiguity given by
\cite{mandel77,jacob77,reutenauer77,ibarra86,weber86}. The most
efficient algorithms for testing polynomial and exponential ambiguity,
and thereby testing finite ambiguity were presented by
\cite{weber87,weber91}. The algorithms presented in \cite{weber91}
assume the input automaton to be $\e$-free, but they are extended to
the case where the automaton has $\e$-transitions in \cite{weber87}.
In the presence of $\e$-transitions, the complexity of the algorithms
given by \cite{weber87} is $O((|A|_E + |A|_Q^2)^2)$ for testing the
exponential ambiguity of an automaton $A$ and $O((|A|_E + |A|_Q^2)^3)$
for testing polynomial ambiguity, where $|A|_E$ stands for the number
of transitions and $|A|_Q$ the number of states of $A$.

This paper presents significantly more efficient algorithms for
testing finite, polynomial, and exponential ambiguity for the general
case of automata with $\e$-transitions. It gives an algorithm for
testing the exponential ambiguity of an automaton $A$ in time
$O(|A|_E^2)$, and finite or polynomial ambiguity in time
$O(|A|_E^3)$. The main idea behind our algorithms is to make use of
the composition or intersection of finite automata with
$\epsilon$-transitions \cite{pereira-riley,ecai}. The $\e$-filter used
in these algorithms crucially helps in the analysis and test of the
ambiguity. We also give an algorithm to determine the degree of
polynomial ambiguity of a finite automaton $A$ that is polynomially
ambiguous in time $O(|A|_E^3)$. Finally, we present an application of
our algorithms to an approximate computation of the entropy of a
probabilistic automaton.

The remainder of the paper is organized as
follows. Section~\ref{sec:preliminaries} presents general automata and
ambiguity definitions. In Section~\ref{sec:characterization} we give a
brief description of existing characterizations for the ambiguity of
automata and extend them to the case of automata with
$\e$-transitions.  In Section~\ref{sec:algorithms} we present our
algorithms for testing the finite, polynomial, and exponential
ambiguity, and the proof of their
correctness. Section~\ref{sec:entropy} details the relevance of these
algorithms to the approximation of the entropy of probabilistic
automata.

\section{Preliminaries}
\label{sec:preliminaries}

\begin{definition}
\label{def:fa}
A {\em finite automaton} $A$ is a 5-tuple $(\Sigma, Q, E, I, F)$
where: $\Sigma$ is a finite alphabet; $Q$ is a finite set of states;
$I \subseteq Q$ the set of initial states; $F \subseteq Q$ the set of
final states; and $E \subseteq Q \times (\Sigma \cup \{\epsilon\})
\times Q$ a finite set of transitions, where $\epsilon$ denotes the
empty string.
\end{definition}

We denote by $|A|_Q$ the number of states, by $|A|_E$ the number of
transitions and by $|A| = |A|_E + |A|_Q$ the size of an automaton $A$.
Given a state $q \in Q$, $E[q]$ denotes the set of transitions leaving
$q$. For two subsets $R \subseteq Q$ and $R' \subseteq Q$, we denote
by $P(R, x, R')$ the set of all paths from a state $q \in R$ to a
state $q' \in R'$ labeled with $x \in \Sigma^*$. We also denote by
$p[\pi]$ the origin state, by $n[\pi]$ the destination state, and by
$i[\pi] \in \Sigma^*$ the label of a path $\pi$.

A string $x \in \Sigma^*$ is accepted by $A$ if it labels a successful
path, {\em i.e.} a path from an initial state to a final state.  A
finite automaton $A$ is {\em trim} if every state of $A$ belongs to a
successful path.  $A$ is {\em unambiguous} if for any string $x \in
\Sigma^*$ there is at most one successful path labeled by $x$ in $A$,
otherwise, $A$ is said {\em ambiguous}.  The {\em degree of ambiguity}
of a string $x$ in $A$, denoted by $\da(A,x)$, is the number of
successful paths in $A$ labeled by $x$. Note that if $A$ contains an
$\e$-cycle, there exist $x \in \Sigma^*$ such that $\da(A,x) =
\infty$. Using a depth-first search restricted to $\e$-transitions, it
can be decided in linear time whether $A$ has $\e$-cycles. Thus, in
the following, we will assume without loss of generality that $A$ is
$\e$-cycle free.

The {\em degree of ambiguity} of $A$ is defined as $\da(A) = \sup_{x
\in \Sigma^*} \da(A,x)$.  $A$ is said {\em finitely ambiguous} if
$\da(A) < \infty$ and {\em infinitely ambiguous} if $\da(A) =
\infty$. $A$ is said {\em polynomially ambiguous} if there exists a
polynomial $h$ in $\Nset[X]$ such that $\da(A,x) \le h(|x|)$ for all
$x \in \Sigma^*$. The minimal degree of such a polynomial is called
the {\em degree of polynomial ambiguity} of $A$, denoted by
$\dpa(A)$. By definition, $\dpa(A) = 0$ iff $A$ is finitely ambiguous.
When $A$ is infinitely ambiguous but not polynomially ambiguous, we
say that $A$ is {\em exponentially ambiguous} and that $\dpa(A) =
\infty$.

\section{Characterization of infinite ambiguity}
\label{sec:characterization}

The characterization and test of finite, polynomial, and exponential
ambiguity of finite automata without $e$-transitions are based on the
following fundamental properties.
\cite{mandel77,jacob77,reutenauer77,ibarra86,weber86,weber87,weber91}.

\begin{figure}[t]
\begin{center}
\begin{tabular}{@{\hspace{2cm}}c@{\hspace{2cm}}c}
\ipsfig{.4}{figure=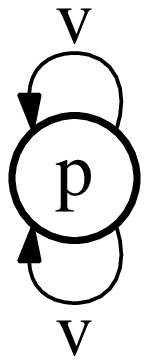} & \ipsfig{.4}{figure=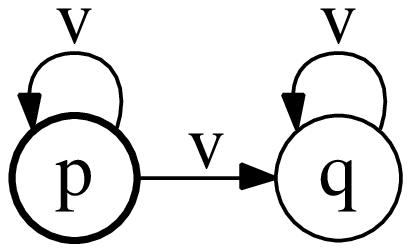}\\
(a) & (b)\\
\multicolumn{2}{c}{\rotatebox{-90}{\ipsfig{.45}{figure=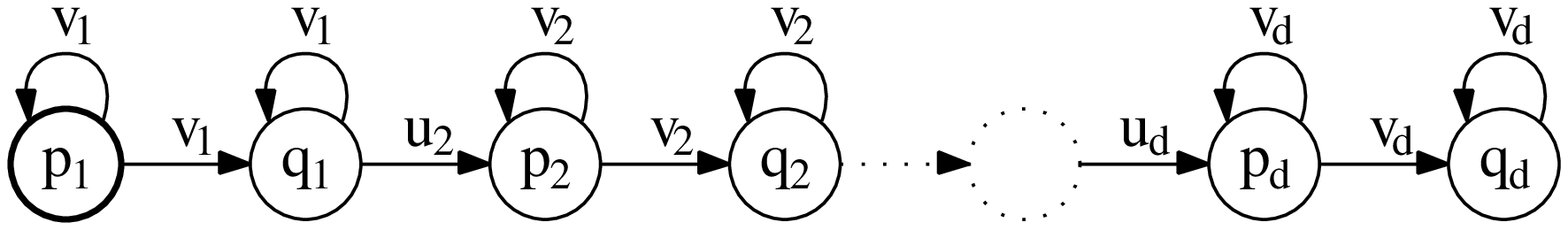}}}\\\\
\multicolumn{2}{c}{(c)}
\end{tabular}
\end{center}
\caption{Illustration of the (a) (EDA), (b) (IDA) and (c) (IDA$_d$) properties.}
\label{fig:ida}
\end{figure}

\begin{definition}
The following are three key properties for the characterization of 
the ambiguity of an automata $A$.
\begin{enumerate}
\item[(a)] {\rm (EDA):} There exists a state $q$ with at least two distinct
  cycles labeled by some $v \in \Sigma^*$ (Figure~\ref{fig:ida}(a)).

\item[(b)] {\rm (IDA):} There exist two distinct states $p$ and $q$ with
  paths labeled with $v$ from $p$ to $p$, $p$ to $q$, and $q$ to $q$,
  for some $v \in \Sigma^*$  (Figure~\ref{fig:ida}(b)).

\item[(c)] {\rm (IDA$_d$):} There exist $2d$ states $p_1, \ldots p_d, q_1,
  \ldots, q_d$ in $A$ and $2d - 1$ strings $v_1, \ldots, v_d$ and
  $u_2, \ldots u_d$ in $\Sigma^*$ such that for all $1 \le i \le d$,
  $p_i\not=q_i$ and $P(p_i, v_i, p_i)$, $P(p_i,v_i,q_i)$ and $P(q_i,
  v_i, q_i)$ are non-empty and for all $2 \le i \le d$, $P(q_{i-1},
  u_i, p_i)$ is non-empty  (Figure~\ref{fig:ida}(c)).
\end{enumerate}
\end{definition}
Observe that (EDA) implies (IDA). Assuming (EDA), let $e$ and $e'$ be
the first transitions that differ in the two cycles at state $q$, then
we must have $n[e] \not= n[e']$ since the definition \ref{def:fa}
disallows multiple transitions between the same two states with the
same label. Thus, (IDA) holds for the pair $(n[e],n[e'])$.

In the $\e$-free case, it was shown that a trim automaton $A$
satisfies (IDA) iff $A$ is infinitely ambiguous
\cite{weber86,weber91}, that $A$ satisfies (EDA) iff $A$ is
exponentially ambiguous \cite{ibarra86}, and that $A$ satisfies
(IDA$_d$) iff $\dpa(A) \ge d$ \cite{weber87,weber91}. These
characterizations can be straightforwardly extended to the case of
automata with $\e$-transitions in the following proposition. 

\begin{proposition}
\label{prop:char}
Let $A$ be a trim $\e$-cycle free finite automaton.
\begin{enumerate}[(i)]
\item $A$ is infinitely
ambiguous iff $A$ satisfies {\rm (IDA)}.
\item $A$ is exponentially ambiguous
iff $A$ satisfies {\rm (EDA)}.
\item $\dpa(A) \ge d$ iff $A$ satisfies {\rm (IDA$_d$)}.
\end{enumerate}
\end{proposition}
\begin{proof}
The proof is by induction on the number of $\e$-transitions
in $A$. If $A$ does not have any $\e$-transitions, 
then the proposition holds as shown in \cite{weber86,weber91}
for (i), \cite{ibarra86} for (ii) and \cite{weber91}
for (iii).

Assume now that $A$ has $n + 1$ $\e$-transitions, $n \ge 0$, and that
the statement of the proposition holds for all automata with $n$
$\e$-transitions. Select an $\e$-transition $e_0$ in $A$, and let $A'$
be the finite automaton obtained after application of $\e$-removal to
$A$ limited to transition $e_0$. $A'$ is obtained by deleting
$e_0$ from $A$ and by adding a transition $(p[e_0], l[e], n[e])$ for
every transition $e \in E[n[e_0]]$. It is clear that $A$ and $A'$ are
equivalent and that there is a label-preserving bijection between the
paths in $A$ and $A'$. Thus, (a) $A$ satisfies (IDA) (resp. (EDA),
(IDA$_d$)) iff $A'$ satisfies (IDA) (resp. (EDA), (IDA$_d$)) and (b)
for all $x \in \Sigma^*$, $\da(A, x) = \da(A', x)$.  By induction,
proposition \ref{prop:char} holds for $A'$ and thus, it follows from
(a) and (b) that proposition \ref{prop:char} also holds for $A$. \qed
\end{proof}
These characterizations have been used in \cite{weber87,weber91} to
design algorithms for testing infinite, polynomial, and exponential
ambiguity, and for computing the degree of polynomial ambiguity in the
$\e$-free case.

\begin{theorem}[\cite{weber87,weber91}]
\label{th:weber-noeps}
Let $A$ be a trim $\e$-free finite automaton.
\begin{enumerate}
\item It is decidable in time $O(|A|_E^3)$ whether $A$ is
infinitely ambiguous.
\item It is decidable in time $O(|A|_E^2)$ whether $A$ is
exponentially ambiguous.
\item The degree of polynomial ambiguity of $A$, $\dpa(A)$,
can be computed in $O(|A|_E^3)$.
\end{enumerate}
\end{theorem}
The first result of theorem \ref{th:weber-noeps} has also been
generalized by \cite{weber87} to the case of automata with
$\e$-transitions but with a significantly worse complexity.

\begin{theorem}[\cite{weber87}] 
Let $A$ be a trim $\e$-cycle free finite automaton.  It is decidable
in time $O((|A|_E + |A|_Q^2)^3)$ whether $A$ is infinitely
ambiguous.
\end{theorem}
The main idea used in \cite{weber87} is to defined from $A$ an
$\e$-free automaton $A'$ such that $A$ is infinitely ambiguous iff
$A'$ is infinitely ambiguous. However, the number of transitions of
$A'$ is $|A|_E + |A|_Q^2$. This explains why the complexity in the
$\e$-transition case is significantly worse than in the $\e$-free
case. A similar approach can be used straightforwardly to test the
exponential ambiguity of $A$ with complexity $O((|A|_E + |A|_Q^2)^2)$
and to compute $\dpa(A)$ when $A$ is polynomially ambiguous with
complexity $O((|A|_E + |A|_Q^2)^3)$.

Note that we give here tighter estimates of the complexity of the
algorithms of \cite{weber87,weber91} where the authors gave
complexities using the loose inequality: $|A|_E \le |\Sigma| \cdot
|A|_Q^2$.

\section{Algorithms}
\label{sec:algorithms}

Our algorithms for testing ambiguity are based on a general algorithm
for the composition or intersection of automata, which we describe in
the following section both to be self-contained, and to give a proof of
the correctness of the $\e$-filter which we have not presented in
earlier publications.

\subsection{Intersection of finite automata}

\begin{figure}[t]
\begin{center}
\begin{tabular*}{\textwidth}{@{\hspace{0cm}}c@{\extracolsep{\fill}}c@{\extracolsep{\fill}}c@{\hspace{0cm}}}
\ipsfig{.35}{figure=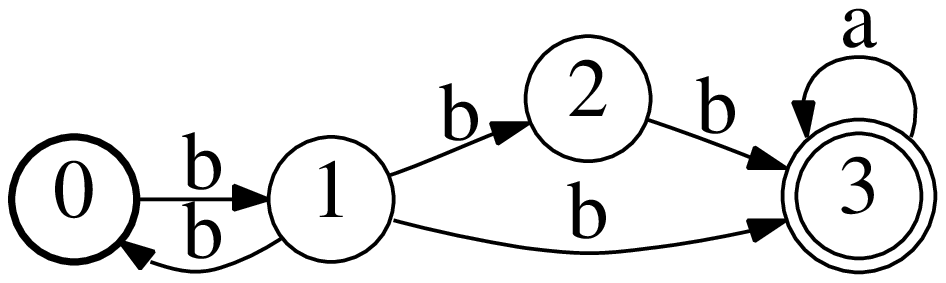} & \ipsfig{.35}{figure=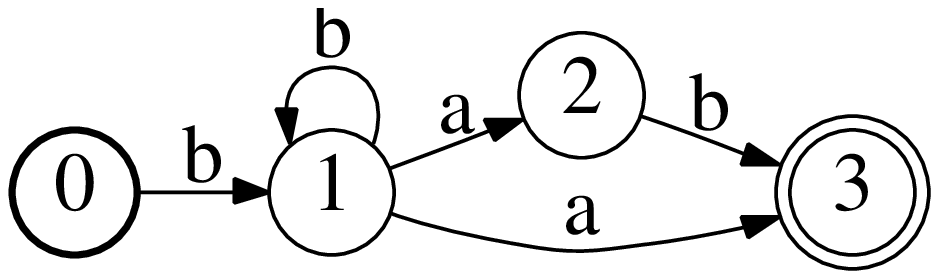} & \raisebox{-.4cm}{\ipsfig{.35}{figure=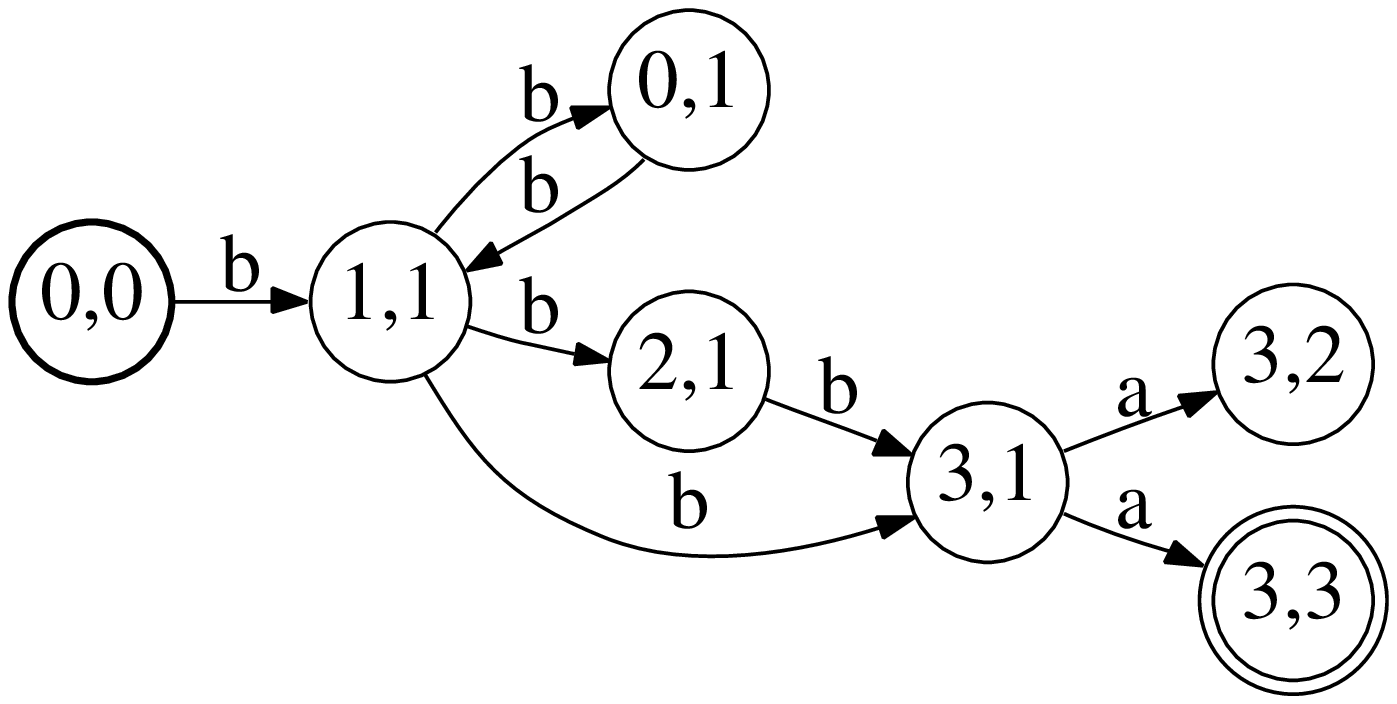}}\\
{\small(a)} & {\small (b)} & {\small(c)}\\
\end{tabular*}
\end{center}
\caption{Example of finite automaton intersection.  (a) Finite automata
$A_1$ and (b) $A_2$. (c) Result of the intersection of $A_1$ and
$A_2$.}
\label{fig:composition}
\end{figure}

The intersection of finite automata is a special case of the general
composition algorithm for weighted transducers
\cite{pereira-riley,ecai}. States in the intersection $A_1 \cap A_2$
of two finite automata $A_1$ and $A_2$ are identified with pairs of a
state of $A_1$ and a state of $A_2$. Leaving aside $\e$-transitions,
the following rule specifies how to compute a transition of $A_1 \cap
A_2$ from appropriate transitions of $A_1$ and $A_2$:
\begin{equation}
(q_1, a, q'_1) \mbox{ and } (q_2, a, q'_2)
\Longrightarrow ((q_1,q'_1), a, (q_2,q'_2)).
\end{equation}
Figure~\ref{fig:composition} illustrates the algorithm. A state
$(q_1,q_2)$ is initial (resp. final) when $q_1$ and $q_2$ are initial
(resp. final).  In the worst case, all transitions of $A_1$ leaving a
state $q_1$ match all those of $A_2$ leaving state $q_2$, thus the
space and time complexity of composition is quadratic: $O(|A_1|
|A_2|)$, or $O(|A_1|_E |A_2|_E)$ when $A_1$ and $A_2$ are trim.

\subsubsection{Epsilon filtering}

A straightforward generalization of the $\e$-free case would generate
redundant $\e$-paths. This is a crucial issue in the more general case
of the intersection of weighted automata over a non-idempotent
semiring, since it would lead to an incorrect result. The weight of
two matching $\e$-paths of the original automata would then be counted
as many times as the number of redundant $\e$-paths generated in the
result, instead of one. It is also a crucial problem in the unweighted
case that we are considering since redundant $\e$-paths can affect the
test of infinite ambiguity, as we shall see in the next section.  A
critical component of the composition algorithm of
\cite{pereira-riley,ecai} consists however of precisely coping with
this problem using a method called {\em epsilon filtering}.

\begin{figure}[t]
\begin{center}
\begin{tabular*}{\textwidth}{@{\extracolsep{\fill}}c@{\extracolsep{\fill}}c@{\extracolsep{\fill}}c@{\extracolsep{\fill}}c@{\extracolsep{\fill}}}
\raisebox{.75cm}{\ipsfig{.45}{figure=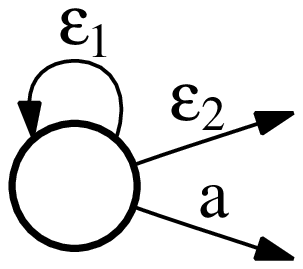}} & 
\raisebox{.75cm}{\ipsfig{.45}{figure=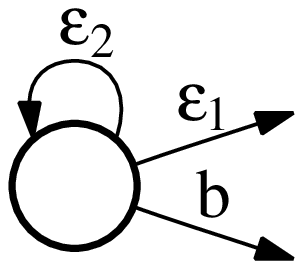}} &
\raisebox{.25cm}{\scalebox{.8}{\newcommand{\radius}{8}
\newcommand{\xa}{0}
\newcommand{\xb}{20}
\newcommand{\xc}{40}

\newcommand{\ya}{40}
\newcommand{\yb}{20}
\newcommand{\yc}{0}

\begin{picture}(45,50)(0,0)

\node[Nw=\radius,Nh=\radius,Nmr=\radius](00)(\xa,\ya){$(0,\!0)$}
\node[Nw=\radius,Nh=\radius,Nmr=\radius](01)(\xa,\yb){$(0,\!1)$}
\node[Nw=\radius,Nh=\radius,Nmr=\radius](02)(\xa,\yc){$(0,\!2)$}

\node[Nw=\radius,Nh=\radius,Nmr=\radius](10)(\xb,\ya){$(1,\!0)$}
\node[Nw=\radius,Nh=\radius,Nmr=\radius](11)(\xb,\yb){$(1,\!1)$}
\node[Nw=\radius,Nh=\radius,Nmr=\radius](12)(\xb,\yc){$(1,\!2)$}

\node[Nw=\radius,Nh=\radius,Nmr=\radius](20)(\xc,\ya){$(2,\!0)$}
\node[Nw=\radius,Nh=\radius,Nmr=\radius](21)(\xc,\yb){$(2,\!1)$}
\node[Nw=\radius,Nh=\radius,Nmr=\radius](22)(\xc,\yc){$(2,\!2)$}

\drawedge[curvedepth=0](00,10){$\e_1{:}\e_1$}
\drawedge[curvedepth=0](01,11){$\e_1{:}\e_1$}
\drawedge[curvedepth=0](02,12){$\e_1{:}\e_1$}

\drawedge[curvedepth=0](10,20){$\e_1{:}\e_1$}
\drawedge[curvedepth=0](11,21){$\e_1{:}\e_1$}
\drawedge[curvedepth=0](12,22){$\e_1{:}\e_1$}

\drawedge[curvedepth=0,ELside=r](00,01){$\e_2{:}\e_2$}
\drawedge[curvedepth=0,ELside=r](01,02){$\e_2{:}\e_2$}

\drawedge[curvedepth=0](10,11){$\e_2{:}\e_2$}
\drawedge[curvedepth=0,linewidth=.4](11,12){$\e_2{:}\e_2$}

\drawedge[curvedepth=0,ELside=l](20,21){$\e_2{:}\e_2$}
\drawedge[curvedepth=0,ELside=l](21,22){$\e_2{:}\e_2$}

\drawedge[curvedepth=0,ELpos=52,linewidth=.4,ELdist=0](00,11){$\e_2{:}\e_1$}
\drawedge[curvedepth=0,ELpos=52,ELdist=0](01,12){$\e_2{:}\e_1$}
\drawedge[curvedepth=0,ELpos=52,ELdist=0](10,21){$\e_2{:}\e_1$}
\drawedge[curvedepth=0,ELpos=52,ELdist=0](11,22){$\e_2{:}\e_1$}

\end{picture} }} & \ipsfig{.5}{figure=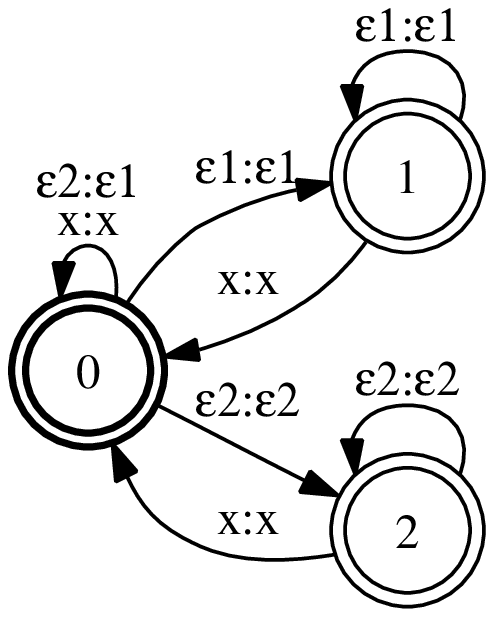}\\
(a) & (b) & (c) & (d)
\end{tabular*}
\end{center}
\caption{Marking of automata, redundant paths and filter. (a)
$\tilde{A}_1$: self-loop labeled with $\e_1$ added at all states of
$A_1$, regular $\e$s renamed to $\e_2$. (b) $\tilde{A}_2$: self-loop
labeled with $\e_2$ added at all states of $A_2$, regular $\e$s
renamed to $\e_1$. (c) Redundant $\e$-paths: a straightforward
generalization of the $\e$-free case could generate all the paths from
$(0, 0)$ to $(2, 2)$ for example, even when composing just two simple
transducers. (d) Filter transducer $M$ allowing a unique $\e$-path.}
\label{fig:e-paths}
\label{fig:marking}
\end{figure}

Figure~\ref{fig:e-paths}(c) illustrates the problem just mentioned.
To match $\e$-paths leaving $q_1$ and those leaving $q_2$, a
generalization of the $\e$-free intersection can make the following
moves: (1) first move forward on an $\e$-transition of $q_1$, or even
a $\e$-path, and stay at the same state $q_2$ in $A_2$, with the hope
of later finding a transition whose label is some label $a \neq \e$
matching a transition of $q_2$ with the same label; (2) proceed
similarly by following an $\e$-transition or $\e$-path leaving $q_2$
while staying at the same state $q_1$ in $A_1$; or, (3) match an
$\e$-transition of $q_1$ with an $\e$-transition of $q_2$.

Let us rename existing $\e$-labels of $A_1$ as $\e_2$, and
existing $\e$-labels of $A_2$ $\e_1$, and let us augment $A_1$
with a self-loop labeled with $\e_1$ at all states and similarly,
augment $A_2$ with a self-loop labeled with $\e_2$ at all states, as
illustrated by Figures~\ref{fig:marking}(a) and (b). These self-loops
correspond to staying at the same state in that machine while
consuming an $\e$-label of the other transition. The three moves just
described now correspond to the matches (1) $(\e_2 \coln \e_2)$,
(2) $(\e_1 \coln \e_1)$, and (3) $(\e_2 \coln \e_1)$. The grid
of Figure~\ref{fig:e-paths}(c) shows all the possible $\e$-paths
between intersection states. We will denote by $\tilde{A}_1$ and
$\tilde{A}_2$ the automata obtained after application of these
changes.

For the result of intersection not to be redundant, between any two of
these states, all but one path must be disallowed.  There are many
possible ways of selecting that path. One natural way is to select the
shortest path with the diagonal transitions ($\e$-matching
transitions) taken first. Figure~\ref{fig:e-paths}(c) illustrates in
boldface the path just described from state $(0, 0)$ to state $(1,
2)$. Remarkably, this filtering mechanism itself can be encoded as a
finite-state transducer such as the transducer $M$ of
Figure~\ref{fig:e-paths}(d). We denote by $(p, q) \preceq (r, s)$ to
indicate that $(r, s)$ can be reached from $(p, q)$ in the grid.

\begin{proposition}
\label{prop:prop1}
Let $M$ be the transducer of Figure~\ref{fig:e-paths}(d). $M$ allows a
unique path between any two states $(p, q)$ and $(r, s)$, with $(p, q)
\preceq (r, s)$.
\end{proposition}

\begin{figure}[t]
\begin{center}
\begin{tabular*}{.8\textwidth}{@{\extracolsep{\fill}}ccc}
\ipsfig{.4}{figure=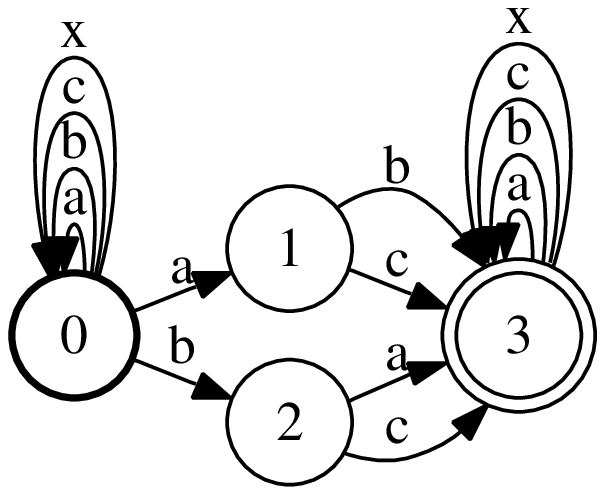}&
\ipsfig{.4}{figure=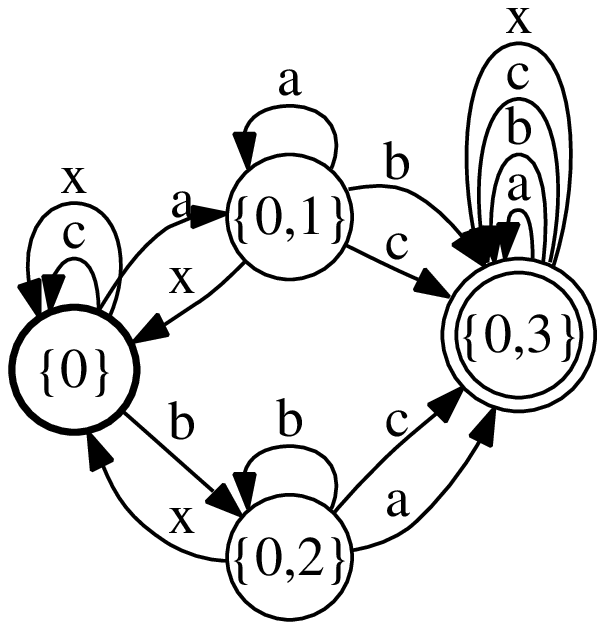}&
\ipsfig{.4}{figure=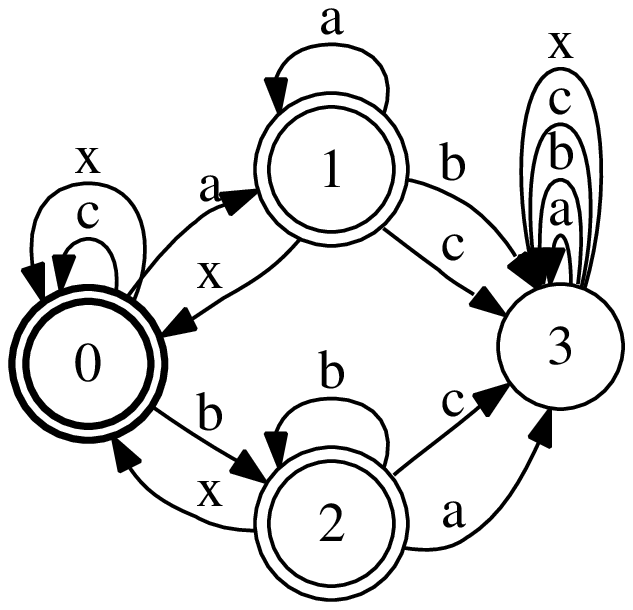}\\
(a) & (b) & (c)
\end{tabular*}
\end{center}
\caption{(a) Finite automaton $A$ representing the set of disallowed
sequences. (b) Automaton $B$, result of the determinization of
$A$. Subsets are indicated at each state. (c) Automaton $C$ obtained
from $B$ by complementation, state $3$ is not coaccessible.}
\label{fig:filter}
\end{figure}

\begin{proof}
Let $a$ denote $(\e_1 \coln \e_1)$, $b$ denote $(\e_2 \coln \e_2)$,
$c$ denote $(\e_2 \coln \e_1)$, and let $x$ stand for any $(x \coln
x)$, with $x \in \Sigma$. The following sequences must be disallowed
by a shortest-path filter with matching transitions first: $ab, ba,
ac, bc$.  This is because, from any state, instead of the moves $ab$
or $ba$, the matching or diagonal transition $c$ can be
taken. Similarly, instead of $ac$ or $bc$, $ca$ and $cb$ can be taken
for an earlier match. Conversely, it is clear from the grid or an
immediate recursion that a filter disallowing these sequences accepts
a unique path between two connected states of the grid.

Let $L$ be the set of sequences over $\sigma = \set{a, b, c, x}$ that
contain one of the disallowed sequence just mentioned as a substring
that is $L = \sigma^* (ab + ba + ac + bc)\sigma^*$. Then
$\overline{L}$ represents exactly the set of paths allowed by that
filter and is thus a regular language. Let $A$ be an automaton
representing $L$ (Figure~\ref{fig:filter}(a)). An automaton
representing $\overline{L}$ can be constructed from $A$ by
determinization and complementation (Figures~\ref{fig:filter}(a)-(c)).
The resulting automaton $C$ is equivalent to the transducer $M$
after removal of the state $3$, which does not admit a path to a final
state.\qed
\end{proof}
Thus, to intersect two finite automata $A_1$ and $A_2$ with
$\e$-transitions, it suffices to compute $\tilde{A}_1 \circ M \circ
\tilde{A}_2$, using the the $\e$-free rules of intersection or
composition.

\begin{theorem}
\label{th:int-paths}
Let $A_1$ and $A_2$ be two finite automata with $\e$-transitions. To
each pair $(\pi_1, \pi_2)$ of successful paths in $A_1$ and $A_2$
sharing the same input label $x \in \Sigma^*$ corresponds a unique
successful path $\pi$ in $A_1 \cap A_2$ labeled by $x$.
\end{theorem}
\begin{proof}
This follows straightforwardly from proposition \ref{prop:prop1}. \qed
\end{proof}

\subsection{Testing for infinite ambiguity}

We start with a test of the exponential ambiguity of $A$. The key is
that the (EDA) property translates into a very simple property for
$A^2 = A \cap A$.

\begin{lemma}
\label{lem:eda-int}
Let $A$ be a trim $\e$-cycle free finite automaton.  $A$ satisfies
{\rm(EDA)} iff there exists a strongly connected component of $A^2 = A
\cap A$ that contains two states of the form $(p,p)$ and $(q,q')$,
where $p$, $q$ and $q'$ are states of $A$ with $q \neq q'$.
\end{lemma}
\begin{proof}
  Assume that $A$ satisfies (EDA). There exist a state $p$ and a
  string $v$ such that there are two distinct cycles $c_1$ and $c_2$
  labeled by $v$ at $p$. Let $e_1$ and $e_2$ be the first edges that
  differ in $c_1$ and $c_2$. We can then write $c_1 = \pi e_1 \pi_1$
  and $c_2 = \pi e_2 \pi_2$. If $e_1$ and $e_2$ share the same label,
  let $\pi'_1 = \pi e_1$, $\pi'_2 = \pi e_2$, $\pi''_1 = \pi_1$ and
  $\pi''_2 = \pi_2$. If $e_1$ and $e_2$ do not share the same label,
  exactly one of them must be an $\e$-transition. By symmetry, we can
  assume without loss of generality that $e_1$ is the $\e$-transition.
  Let $\pi'_1= \pi e_1$, $\pi'_2 = \pi$, $\pi''_1 = \pi_1$ and
  $\pi''_2 = \e_2 \pi_2$.  In both cases, let $q = n[\pi'_1] =
  p[\pi''_1]$ and $q'= n[\pi'_2] = p[\pi''_2]$.  Observe that $q \not=
  q'$. Since $i[\pi'_1] = i[\pi'_2]$, $\pi'_1$ and $\pi'_2$ are
  matched by intersection resulting in a path in $A^2$ from $(p,p)$ to
  $(q,q')$. Similarly, since $i[\pi''_1] = i[\pi''_2]$, $\pi''_1$ and
  $\pi''_2$ are matched by intersection resulting in a path from
  $(q,q')$ to $(p,p)$. Thus, $(p,p)$ and $(q,q')$ are in the same
  strongly connected component of $A^2$.

  Conversely, assume that there exist states $p$, $q$ and $q'$ in $A$
  such that $q \not= q'$ and that $(p,p)$ and $(q,q')$ are in the same
  strongly connected component of $A^2$. Let $c$ be a cycle in $(p,p)$
  going through $(q,q')$, it has been obtained by matching two cycles
  $c_1$ and $c_2$. If $c_1$ were equal to $c_2$, intersection would
  match these two paths creating a path $c'$ along which all the
  states would be of the form $(r,r)$, and since $A$ is trim this
  would contradict Theorem~\ref{th:int-paths}. Thus, $c_1$ and $c_2$
  are distinct and (EDA) holds. \qed
\end{proof}
Lemma \ref{lem:eda-int} leads to a straightforward algorithm for
testing exponential ambiguity.

\begin{theorem}
\label{th:eda-int}
Let $A$ be a trim $\e$-cycle free finite automaton. It is decidable
in time $O(|A|_E^2)$ whether $A$ is exponentially ambiguous.
\end{theorem}
\begin{proof}
  The algorithm proceeds as follows. We compute $A^2$ and, using a
  depth-first search of $A^2$, trim it and compute its strongly
  connected components. It follows from Lemma~\ref{lem:eda-int} that
  $A$ is exponentially ambiguous iff there is a strongly connected
  component that contains two states of the form $(p,p)$ and $(q,q')$
  with $q \not= q'$. Finding such a strongly connected component can
  be done in time linear in the size of $A^2$, {\em i.e.} in
  $O(|A|_E^2)$ since $A$ and $A^2$ are trim. Thus, the complexity of
  the algorithm is in $O(|A_E|^2)$. \qed
\end{proof}
Testing the (IDA) property requires finding three paths sharing the
same label in $A$. This can be done in a natural way using the
automaton $A^3 = A \cap A \cap A$, as shown below.

\begin{lemma}
\label{lem:ida-int}
Let $A$ be a trim $\e$-cycle free finite automaton.  $A$ satisfies
{\rm(IDA)} iff there exist two distinct states $p$ and $q$ in $A$ with
a non-$\e$ path in $A^3 = A \cap A \cap A$ from state $(p,p,q)$ to
state $(p,q,q)$.
\end{lemma}
\begin{proof}
  Assume that $A$ satisfies (IDA). Then, there exists a string $v \in
  \Sigma^*$ with three paths $\pi_1 \in P(p,v,p)$, $\pi_2 \in
  P(p,v,q)$ and $\pi_3 \in P(q,v,p)$. Since these three paths share
  the same label $v$, they are matched by intersection resulting in a
  path $\pi$ in $A^3$ labeled with $v$ from $(p[\pi_1], p[\pi_2],
  p[\pi_3]) = (p,p,q)$ to $(n[\pi_1], n[\pi_2], n[\pi_3]) = (p, q,
  q)$.

  Conversely, if there is a non-$\e$ path $\pi$ form $(p,p,q)$ to
  $(p,q,q)$ in $A^3$, it has been obtained by matching three paths
  $\pi_1$, $\pi_2$ and $\pi_3$ in $A$ with the same input $v = i[\pi]
  \not= \e$. Thus, (IDA) holds. \qed
\end{proof}
Finally, Theorem~\ref{th:eda-int} and Lemma~\ref{lem:ida-int} can be
combined to yield the following result.

\begin{theorem}
\label{th:ida-int}
Let $A$ be a trim $\e$-cycle free finite automaton.  It is decidable
in time $O(|A|_E^3)$ whether $A$ is finitely, polynomially, or
exponentially ambiguous.
\end{theorem}
\begin{proof}
  First, Theorem~\ref{th:eda-int} can be used to test whether $A$ is
  exponentially ambiguous by computing $A^2$. The complexity of this
  step is $O(|A|_E^2)$.

  If $A$ is not exponentially ambiguous, we proceed by computing and
  trimming $A^3$ and then testing whether $A^3$ verifies the property
  described in lemma \ref{lem:ida-int}. This is done by considering
  the automaton $B$ on the alphabet $\Sigma' = \Sigma \cup \set{\#}$
  obtained from $A^3$ by adding a transition labeled by $\#$ from
  state $(p,q,q)$ to state $(p,p,q)$ for every pair $(p, q)$ of states
  in $A$ such that $p \not= q$. It follows that $A^3$ verifies the
  condition in lemma \ref{lem:ida-int} iff there is a cycle in $B$
  containing both a transition labeled by $\#$ and a transition
  labeled by a symbol in $\Sigma$.  This property can be checked
  straightforwardly using a depth-first search of $B$ to compute its
  strongly connected components. If a strongly connected component of
  $B$ is found that contains both a transition labeled with $\#$ and a
  transition labeled by a symbol in $\Sigma$, $A$ verifies (IDA) but
  not (EDA) and thus $A$ is polynomially ambiguous. Otherwise, $A$ is
  finitely ambiguous. The complexity of this step is linear in the
  size of $B$: $O(|B|_E) = O(|A_E|^3 + |A_Q|^2) = O(|A_E|^3)$ since
  $A$ and $B$ are trim.

  The total complexity of the algorithm is $O(|A|_E^2 + |A|_E^3) =
  O(|A|_E^3)$.
\end{proof}
When $A$ is polynomially ambiguous, we can derive from the
algorithm just described one that computes $\dpa(A)$.

\begin{theorem}
\label{th:deg}
Let $A$ be a trim $\e$-cycle free finite automaton. If $A$ is polynomially
ambiguous, $\dpa(A)$ can be computed in time $O(|A|_E^3)$.
\end{theorem}
\begin{proof}
  We first compute $A^3$ and use the algorithm of theorem
  \ref{th:ida-int} to test whether $A$ is polynomially ambiguous and
  to compute all the pairs $(p,q)$ that verify the condition of
  Lemma~\ref{lem:ida-int}. This step has complexity $O(|A|_E^3)$.

  We then compute the component graph $G$ of $A$, and for each pair
  $(p,q)$ found in the previous step, we add a transition labeled with
  $\#$ from the strongly connected component of $p$ to the one of $q$.
  If there is a path in that graph containing $d$ edges labeled by
  $\#$, then $A$ verifies (IDA$_d$). Thus, $\dpa(A)$ is the maximum
  number of edges marked by $\#$ that can be found along a path in
  $G$. Since $G$ is acyclic, this number can be computed in linear
  time in the size of $G$, {\em i.e.} in $O(|A|_Q^2)$. Thus, the
  overall complexity of the algorithm is $O(|A|_E^3)$. \qed
\end{proof}

\section{Application to the Approximation of Entropy}
\label{sec:entropy}

In this section, we describe an application in which determining the
degree of ambiguity of a {\em probabilistic} automaton helps estimate
the quality of an approximation of its entropy.

Weighted automata are automata in which each transition carries some
weight in addition to the usual alphabet symbol. The weights are
elements of a semiring, that is a ring that may lack negation. The following
is a more formal definition.

\begin{definition}
A {\em weighted automaton} $A$ over a semiring $\K$ is a 7-tuple
$(\Sigma, Q, I, F, E, \lambda, \rho)$ where: $\Sigma$ is the finite
alphabet of the automaton, $Q$ is a finite set of states, $I \subseteq
Q$ the set of initial states, $F \subseteq Q$ the set of final states,
$E \subseteq Q \times \Sigma \cup \set{\epsilon} \times \Kset \times
Q$ a finite set of transitions, $\lambda: I \rightarrow \Kset$ the
initial weight function mapping $I$ to $\Kset$, and $\rho: F
\rightarrow \Kset$ the final weight function mapping $F$ to $\Kset$.
\end{definition}
Given a transition $e \in E$, we denote by $w[e]$ its weight. We
extend the weight function $w$ to paths by defining the weight of a
path as the $\otimes$-product of the weights of its constituent
transitions: $w[\pi] = w[e_1] \otimes \cdots \otimes w[e_k]$. The
weight associated by a weighted automaton $A$ to an input string $x
\in \Sigma^*$ is defined by:
\begin{equation}
\den{A}(x) = \bigoplus_{\pi \in P(I, x, F)} \lambda[p[\pi]]
\otimes w[\pi] \otimes \rho[n[\pi]].
\end{equation}
The entropy $H(A)$ of a probabilistic automaton $A$ is defined as:
\begin{equation}
  H(A) = - \sum_{x \in \Sigma^*} \den{A}(x) \log (\den{A}(x)).
\end{equation}
Let $\Kset$ denote $(\Rset \cup \{+\infty, -\infty\}) \times (\Rset
\cup \{+\infty, -\infty\})$.  The system $(\Kset, \oplus, \otimes, (0,
0), (1, 0))$ where $\oplus$ and $\otimes$ are defined as follows
defines a commutative semiring called the \emph{entropy semiring}
\cite{cmrr}. For any two pairs $(x_1, y_1)$ and $(x_2, y_2)$ in
$\Kset$,
\begin{eqnarray}
(x_1, y_1) \oplus (x_2, y_2) & = & (x_1 + x_2, y_1 + y_2)\\
(x_1, y_1) \otimes (x_2, y_2) & = & (x_1 x_2, x_1 y_2 + x_2 y_1).
\end{eqnarray}
In \cite{cmrr}, the authors show that a generalized shortest-distance
algorithm over this semiring correctly computes the entropy of an
unambiguous probabilistic automaton $A$. The algorithm starts by
mapping the weight of each transition to a pair where the first
element is the probability and the second the entropy: $w[e] \mapsto
(w[e], - w[e] \log w[e])$. The algorithm then proceeds by computing
the generalized shortest-distance under the {\em entropy semiring},
which computes the $\oplus$-sum of the weights of all accepting paths
in $A$.

In this section, we show that the same shortest-distance algorithm
yields an approximation of the entropy of an ambiguous probabilistic
automaton $A$, where the approximation quality is a function of the
degree of polynomial ambiguity, $\dpa({A})$. Our proofs make use of
the standard log-sum inequality \cite{cover}, a special case of
Jensen's inequality, which holds for any positive reals $a_1, \ldots,
a_k$, and $b_1, \ldots, b_k$:
\begin{equation}
\label{eqn:logsum}
\sum_{i=1}^k a_i \log \frac{a_i}{b_i} \geq \left(\sum_{i=1}^k
a_i\right) \log \frac{\sum_{i=1}^k a_i}{\sum_{i=1}^k b_i}.
\end{equation}

\begin{lemma}
\label{lemma:log-sum}
Let $A$ be a probabilistic automaton and let $x \in \Sigma^+$ be a
string accepted by $A$ on $k$ paths $\pi_1, \ldots, \pi_k$. Let
$w(\pi_i)$ be the probability of path $\pi_i$. Clearly, $\den{A}(x) =
\sum_{i=1}^k w(\pi_i)$. Then,
\begin{equation}
  \sum_{i=1}^k w(\pi_i) \log w(\pi_i) \geq \den{A}(x) ( \log
  \den{A}(x) - \log{k} ).
\end{equation}
\end{lemma}
\begin{proof}
  The result follows straightforwardly from the log-sum inequality, with $a_i =
  w(\pi_i)$ and $b_i = 1$:
  \begin{equation}
\small    \sum_{i=1}^{k} w(\pi_i) \log w(\pi_i) \geq \left(\sum_{i=1}^{k}
    w(\pi_i)\right) \log \frac{\sum_{i=1}^{k} w(\pi_i)}{k} = \den{A}(x) (\log \den{A}(x) - \log k).
  \end{equation}\qed
\end{proof}
For a probabilistic automaton $A$, let $S(A)$ be the quantity computed
by the generalized shortest-distance algorithm with the entropy
semiring. For an unambiguous automaton $A$, $S(A) = H(A)$ \cite{cmrr}.

\begin{theorem}
Let $A$ be a probabilistic automaton and let $L$ denote the expected
length of strings accepted by $A$ (i.e.~$L = \sum_{x \in \Sigma^*} |x|
\den{A}(x)$). Then, 
\begin{enumerate}
  \item{If $A$ is {\em finitely ambiguous} with degree of ambiguity $k$
  (i.e.~$\da(A) = k$ for some $k \in \Nset$), then $H(A) \leq S(A) \leq H(A) +
  \log k$.}
  \item{If $A$ is {\em polynomially ambiguous} with degree of
  polynomial ambiguity $k$ (i.e.~$\dpa(A) = k$ for some $k \in \Nset$),
  then $H(A) \leq S(A) \leq H(A) + k \log L$.}
\end{enumerate}
\end{theorem}
\begin{proof}
  The lower bound, $S(A) \geq H(A)$ follows from the observation that
  for a string $x$ that is accepted in $A$ by $k$ paths $\pi_1,
  \ldots, \pi_k$,
  \begin{equation}
    \sum_{i=1}^k w(\pi_i) \log (w (\pi_i)) \leq (\sum_{i=1}^k
    w(\pi_i)) \log ( \sum_{i=1}^k w(\pi_i)).
  \end{equation}
  Since the quantity $-\sum_{i=1}^k w(\pi_i) \log (w (\pi_i))$ is
  string $x$'s contribution to $S(A)$ and the quantity $-(\sum_{i=1}^k
  w(\pi_i)) \log ( \sum_{i=1}^k w(\pi_i))$ its contribution to $H(A)$,
  summing over all accepted strings $x$, we obtain $H(A) \leq S(A)$.

  Assume that $A$ is finitely ambiguous with degree of ambiguity $k$. Let $x
  \in \Sigma^*$ be a string that is accepted on $l_x \leq k$ paths
  $\pi_1, \ldots, \pi_{l_x}$. By Lemma~\ref{lemma:log-sum},
  \begin{equation}
  \small  \sum_{i=1}^{l_x} w(\pi_i) \log w(\pi_i) \geq \den{A}(x) ( \log
    \den{A}(x) - \log l_x ) \geq \den{A}(x) ( \log \den{A}(x) -
    \log k ).
  \end{equation}
  Thus, 
  \begin{equation}
\small    S(A) = - \sum_{x \in \Sigma^*} \sum_{i=1}^{l_x} w(\pi_i) \log
    w(\pi_i) \leq H(A) + \sum_{x \in \Sigma^*} (\log k) \den{A}(x) = H(A) + \log k.
  \end{equation}
  This proves the first statement of the theorem.  

  Next, assume that $A$ is polynomially ambiguous with degree of
  polynomial ambiguity $k$. By Lemma~\ref{lemma:log-sum},
  \begin{equation}
    \small \sum_{i=1}^{l_x} w(\pi_i) \log w(\pi_i) \geq \den{A}(x) ( \log
    \den{A}(x) - \log l_x ) \geq \den{A}(x) ( \log \den{A}(x) -
    \log (|x|^k) ).
  \end{equation}
  Thus,
  \begin{eqnarray}
    S(A) & \leq &  H(A) + \sum_{x \in \Sigma^*} k  \den{A}(x) \log
     |x| = H(A) + k \mathbb{E}_A[\log |x|]\\
     & \leq & H(A) + k \log \mathbb{E}_A[|x|] = H(A) + k \log L, \qquad (\mbox{by
       Jensen's inequality}) \nonumber
  \end{eqnarray}
which proves the second statement of the theorem. \qed
\end{proof}
The quality of the approximation of the entropy of a probabilistic
automaton $A$ depends on the expected length $L$ of an accepted
string. $L$ can be computed efficiently for an arbitrary probabilistic
automaton using the {\em expectation semiring} and the generalized
shortest-distance algorithms, using techniques similar to the ones
described in \cite{cmrr}. The definition of the expectation semiring
is identical to the entropy semiring. The only difference is in the
initial step, where the weight of each transition in $A$ is mapped to
a pair of elements. Under the expectation semiring, the mapping is
$w[e] \mapsto (w[e], w[e])$.

\section{Conclusion}
\label{sec:conclusion}

We presented simple and efficient algorithms for testing the finite,
polynomial, or exponential ambiguity of finite automata with
$\e$-transitions. We conjecture that the running-time complexity of
our algorithms is optimal. These algorithms have a variety of
applications, in particular to test a pre-condition for the
applicability of other automata algorithms. Our application to the
approximation of the entropy gives another illustration of 
the applications of these algorithms.

Our algorithms also illustrate the prominent role played by the
general algorithm for the intersection or composition of automata and
transducers with $\e$-transitions in the design of \emph{testing
  algorithms}. Composition can be used to devise simple and efficient
testing algorithms. We have shown elsewhere how it can be used to test
the functionality of a finite-state transducer or to test the twins
property for weighted automata and transducers \cite{twins}.

\subsubsection*{Acknowledgments.}
{\footnotesize The research of Cyril Allauzen and Mehryar Mohri was
partially supported by the New York State Office of Science Technology
and Academic Research (NYSTAR). This project was also sponsored in
part by the Department of the Army Award Number W81XWH-04-1-0307. The
U.S. Army Medical Research Acquisition Activity, 820 Chandler Street,
Fort Detrick MD 21702-5014 is the awarding and administering
acquisition office.  The content of this material does not necessarily
reflect the position or the policy of the Government and no official
endorsement should be inferred.}

\bibliographystyle{plain}
\bibliography{amb}
\end{document}